**Protein Folding/Unfolding Phenomenon is originated by Synchronization/Desynchronization of Instantaneously Induced Oscillating Dipoles. A Hypothesis.**

Germán Miño-Galaz, PhD.

Departamento de Ciencias Físicas, Facultad de Ciencias Exactas, Universidad Andres Bello, Republica 498, Santiago, Chile
germ.mino@uandresbello.edu
german.mino.galaz@gmail.com

**Abstract**

Protein folding is a phenomenon that has been studied for about 50 years and still remains as an unsolved problem. The main feature of this process is that it occurs as an all or none process, so a protein, can jump directly between folded and unfolded states. It is proposed in this manuscript that the cooperative phenomena associated to protein folding has its origin in the synchronization oscillating phases of the instantaneously induced dipoles that gives rise to the van der Waals London dispersion interaction. When the oscillation of the induced dipoles is synchronized an enhanced interaction regime is triggered and the protein enters a into a folding process. The propagation of this regime throughout the molecular structure complete the folding process. If a desynchronizing strong enough perturbation is introduced at any component cooperative network the system enters in a weakened -or even a repulsive- van der Waals dispersion interaction regime that propagate itself throughout the structure, that triggers a phase transition that finally unfolds the protein.

**The protein folding problem**

Protein folding is a phenomenon that has been studied for about 50 years. Despite the theoretical (Voelz 2012, Piana 2013, Kuhlman 2019, Jafari 2020, Noe 2020, Senior 2020, Tavernelli 2020) and the experimental (Dill 2007, Xu 2006,Wang 2004, Voelz 2012, Kuhlman 2019) efforts to solve it, still remains as an unsolved problem (Dill 1985, Dill 1990, Dill 1993, Dill 2007, Dill 2012, Voelz 2012, Kuhlman 2019). Protein folding is a cooperative process that depending on the specific sequence and size of the protein takes from nanoseconds to milliseconds to complete. Cooperativity in protein folding is generated by an immense set of small interactions, operating in consort, so if one or a small group of those interactions is perturbed the whole network is affected. Nevertheless, the precise object underpinning this cooperativity still remains unidentified. Cooperativity obliterates the exploration of the full configurational space available to the unfolded protein, allowing an astonishingly fast convergence to its correct folded state. The Levinthal paradox clearly illuminates this. For example, for a small size protein of 101 amino acids, with each amino acid sampling 3 separated conformations, the number of possible configurations is $3^{100} = 5 \times 10^{47}$. (Zwanzig 1991) Even if the protein is able to sample a new configuration at the rate of $10^{13}$ Hz, it will take $10^{27}$ years to try all the possible configurations. The main feature of this process is that it occurs as an all or none process, so a protein, in aqueous solution, can jump directly between folded and unfolded states. Those states depend on external factors such as temperature, pH or the concentration of denaturing agents.

**Dispersion Interaction.**

Many intermolecular and intramolecular interactions mediate the occurrence of protein folding (Dill 1993). The main one among them is the dispersion interaction. As has been shown by molecular simulation the folded state of proteins is highly dependent of the dispersion interaction, so when weakened or omitted an unfolded or disordered state manifests. On one hand, using classical molecular simulations, Cerny have analyzed the change from 1 to 0.01 the ε constant of the Lennard-Jones potential as a proxy for the modulation of the dispersion interaction. So, when the ε value is changed from 0.5 or to 0.01 it leads to the loss of the tridimensional structure of simulated models of proteins (Cerny 2009). On the other hand, by quantum molecular simulation Vondrásek has shown the relevant role of dispersion interaction in the stability of the hydrophobic core of proteins, which is characteristically composed by hydrophobic amino acids such as valine, leucine, isoleucine, phenylalanine and tryptophan (Vondrásek 2005, Vondrásek 2007, Berka 2009). In the same line, Danovich has shown that the C-H fragment of hydrophobic chemical moieties have the possibility of forming dihydrogen bonds C-H … H-C (Danovich 2013, Echeverria, 2011). Those bonds, termed sticky fingers, are stabilized by valence bond (VB) structures of the form $C^+H^-…H^+C^-$ and $C^-H^+…H^-C^+$ (Danovich 2013). Those VB structures are consistent with the classical mechanism of oscillating dipoles as the source of dispersion interaction by the generation of electrostatic interaction that stabilize the dihydrogen bond. Dispersion interaction may be understood as originated by oscillating electric dipoles. Those electric dipoles emerge from charge fluctuation of the electron distribution (London 1930, Voet 1995, Rodrigues 2011) that compose every segment of an atom or molecule and hence of proteins. The model of oscillating dipoles comes after Drude (Drude 1900), and its implementation in classical or quantum molecular simulation, have improved the description of the aggregation states of the objects in both classical (Huang 2014) and quantum simulation (Kleshchonok 2018). In this direction, it has been shown that the folding/unfolding process of small helical peptides acquire a better description from classical molecular models that include Drude oscillators in its calculation algorithms (Huang, 2014). Although the heuristic understanding of the classical models of oscillating dipoles for the dispersion interaction is criticized (Hirschfelder, 1964), the quantum models of oscillating dipoles are outstanding for the description of the dispersion interaction among molecules (Kleshchonok 2018, Jones 2013). At this point we refer to the Quantum Drude Oscillator (QDO) model (Jones 2013). This model is a coarse-grain version molecular quantum mechanics that replace the fine structure of each atom by an oscillator composed by a negatively charged particle -a drudon- that orbits the Drude nucleus. Each quantum Drude oscillator is characterized by a mass, frequency and charge, and interacts with other oscillators by a long-range Coulomb force. So, at an inter-molecular scale, a proper model, although not unique for the description of dispersion interactions, is the QDO model (Jones 2013, Kleshchonok 2018).

**Kuramoto´s oscillators**

Mathematical models to tackle the problem of the collective synchronization of a population of coupled oscillators have been developed (Winfree 1967, Kuramoto 1989). The models have been applied to different fields such as economics, social behavior, cricket chirping, neural networks, Bose-Einstein condensates, among other phenomenon (Strogatz 2000, Acebrón 2005, Arenas 2008). In those models each oscillator exerts a phase dependent influence on the others, and the modulation of their oscillatory rhythm is then allowed. When the natural frequency of the oscillators is too diverse compared to the strength of the coupling, the oscillators are unable to synchronize and the whole system behaves incoherently. However, if the coupling is strong enough, the population of oscillators enters a synchronized coherent state. This transition occurs at the "onset of synchronization" (Winfree 1967, Kuramoto 1989) and is analogous to a phase transition (Daido 1990, Acebrón 2005, Arenas 2008, Strogatz 2000, Pikovsky 2002). The direct similitude between the experimental plots for protein folding/unfolding as function of temperature or by the addition of denaturation agents (Weiss 2000, Volkening 2019), with the plots to the phase transmission of Kuramoto models (Arenas 2008) suggest the existence of oscillators of modulable coupling in protein structure that when synchronized leads to protein folding. When those oscillators are desynchronized the denaturation or unfolding of the protein structure manifest.

**Dispersion interaction is not a constant**

Dispersion interaction is modeled, in classical molecular dynamics of proteins, using the Lennard-Jones potential, that uses a long-range attractive tail with a $D^{-6}$ dependence, where D is the distance of two or more atomic centers. Nevertheless, it has been clearly remarked in physicochemical (Mahanty 1975, Feiler 2008) and quantum studies (Dobson 2015, Ambrosetti 2016) that the van der Waals dispersion interaction is variable. In other words, dispersion interaction depends of the distance D and the number of particles N involved (Dobson 2015). Furthermore, a slight change in this power law may have a profound impact in the interaction properties of the materials (Ambrosetti 2016). On the theoretical side, quantum studies have demonstrated a departure from this Lennard-Jones regime in the modeling of the dispersion interaction as function of distance D between carbine wires, graphene sheets, carbon nanotube and carbyne wire – protein models (Ambrosetti 2016). Other evidence however, has shown that the exchange between $D^{-6}$ to the $D^{-7}$ regimes of the van der Waals dispersion interaction determines the state of stiction of a microelectromechanical systems (Rodriguez 2011). Also, a repulsive version of the van der Waals interaction (Feiler 2008) has been reported. The transition from the D-6 to $D^{-7}$ regime is due to the finite velocity of light that induce a distance dependent relativistic retardation effect (Mahanty, 1975. Dzyaloshinskii 1992. Israelachvili, 2011, Rodriguez 2011). When an instantaneously oscillating dipole is generated it will polarize its neighborhood inducing other transient oscillating dipoles. At short distances the interaction between both oscillating dipoles will be essentially instantaneous. Nevertheless, with the increase of distance between the dipoles the actualization of the electromagnetic signal may find a second dipole out of phase, leading to a weakened attraction that alters the usual $D^{-6}$ regime associated to the dispersion interaction. In particular, experimental work on two curved mica surfaces has shown that the transition from the attractive to retarded regimes in water and other solvents has a c.a. 50 Å limit (Israelachvili, 2011). Those facts open the possibility to postulate the existence of a mechanism that, by changes in the van der Waals dispersion interaction, controls the phenomenology of protein folding/unfolding. Taking into consideration that the full extension of a given protein such as the dimeric Trp-repressor, a 2.4 % of extension of a total of 850 Å (Miño 2013) would be needed to reach the 50 Å limit of the retarded $D^{-7}$ regime. Moreover, a further extension of the protein will increase the protein elements, namely, amino acid residues and backbone components, that enter to the 50 Å limit collaborating with the retarded or repulsive regimes. According to this analysis it is possible to think that when given amino acids residues of a protein that enter in the $D^{-6}$ attractive regime the system becomes sticky or adhesive, and when a protein enters to the retarded $D^{-7}$ regime the same segments or moieties of the protein loose its stickiness and the whole system starts to unfold. Thus, in protein folding phenomenon the alternation between attractive, retarded or even repulsive versions of the van der Waals dispersion interaction in the determination of the folded or unfolded state of proteins seems fairly plausible.

**The Hypothesis**

The concept of synchronization and desynchronization of the origin of instantaneously induced London dipoles interaction is loosely mentioned by some textbooks on the experimental field of surface and interface adhesion (Wu 1982, Pugh 1993, Stokes 1996). Also, a suggestion for this idea arouse from Weinhold : "… retardation allows the dipole fluctuation to ´wooble´ out of perfect alignment, weakening the dispersion interaction…" (Weinhold 2005). Following this line of thought, it is actually apparent that the synchronization of the oscillating phases leads to the entrance in the $D^{-6}$ regime while the desynchronization of the phases leads to the retarded or even to repulsive regimes in molecular structure. Thus, it is proposed in this manuscript that the cooperative phenomena associated to protein folding has its origin in the oscillating phases of the instantaneously induced dipoles. When the oscillation of the induced dipoles is synchronized an enhanced interacting regime is triggered, some components become sticky, and the protein structure start to fold. The propagation of this regime throughout the molecular structure complete the folding process. If a perturbation is introduced in any component of the synchronized network it has the possibility to affect the whole network. If this the perturbation is strong enough the cooperative network is significantly affected and enters in a weakened or repulsive regime that propagate itself throughout the structure. In this hypothesis, this weakened or repulsive regime is associated to the desynchronization of the oscillating network and triggers a phase transition that finally unfolds the protein.

**Some Other Estimates**

From the report QDO model (Jones 2013, Table I) it is possible to obtain the frequency values of Drude oscillators for a set of atoms to tetratomic molecules. Those values lie between 0.06 to 1.3 atomic units which are equivalent to wave length of 6 to 120 nm. Those values fall into the Far UV range. It is well known that proteins are denatured in UV light in this range of frequency (Hussain 2018). Proposed mechanisms for the unfolding of proteins in front of vacuum and extreme UV involves radical formation (Wallace 2001) and thermal effect (Miles 2008). In the context of the proposed hypothesis it is possible that the UV radiation is directly exiting the electronic oscillatory modes of the instantaneously induced dipoles that gives rise to the van der Waals dispersion interaction acting through a non-thermal effect. This excitation may affect the network of coupled oscillators leading to the instabilities of the network, and finally lead to protein denaturation. Thus, the proposed hypothesis may be ascribed as a new possible mechanism that is mediating the UV-induced protein denaturation phenomenon.

**Some paths to validate hypothesis**

Possibilities for research may include, among other approaches, (i) the integration of the Hydrophobic-polar protein folding model (Dill 1985) with algorithms of Kuramoto oscillators.; (ii) the development operators in the context of QDO models (Jones 2013) to extract synchronization/desynchronization information about the interacting molecules or chemical moieties and (iii) the experimental bombardment with specific UV lasers of proteins in bulk or *in singulo*. The latter may use an optical tweezers setup, that combined with specific controls of radical formation and temperature control may discard those effects. Although this last approach is still unclear, it may open the road to experimentally explore the proposed mechanism of protein folding/unfolding mediated by synchronization/desynchronization of oscillatory electric dipoles.